\documentclass[aps,prd,amsmath,superscriptaddress,preprintnumbers,amssymb,floatfix,nofootinbib,longbibliography,twocolumn, ]{revtex4-2} 
\pdfoutput=1

\usepackage{graphicx}
\usepackage{bm}
\usepackage{hyperref}
\usepackage{slashed}
\usepackage{booktabs}

\usepackage{lipsum}
\usepackage[caption=false]{subfig}
\usepackage{multirow}
\usepackage[normalem]{ulem}

\hypersetup{
     colorlinks   = true,
     citecolor    = magenta,
     urlcolor     = blue,
     linkcolor    = magenta
}

\usepackage{listings}
\usepackage[dvipsnames]{xcolor}

\begin{document}

\title{Natural Chlorine Enhances Antineutrino Detection in the Deep Ocean}

\author{Ruike Cao}
\email{ruike\_cao@sjtu.edu.cn}
\affiliation{State Key Laboratory of Dark Matter Physics, Tsung-Dao Lee Institute, Shanghai Jiao Tong University, Shanghai 201210, China}

\author{Iwan Morton-Blake}
\email{Corresponding author, iblake@sjtu.edu.cn} 
\affiliation{State Key Laboratory of Dark Matter Physics, Tsung-Dao Lee Institute, Shanghai Jiao Tong University, Shanghai 201210, China}

\author{Donglian Xu}
\email{Corresponding author, donglianxu@sjtu.edu.cn}
\affiliation{State Key Laboratory of Dark Matter Physics, Tsung-Dao Lee Institute, Shanghai Jiao Tong University, Shanghai 201210, China}
\affiliation{Key Laboratory for Particle Astrophysics and Cosmology (MOE) \& Shanghai Key Laboratory for Particle Physics and Cosmology, Shanghai Jiao Tong University, Shanghai 200240, China}
\affiliation{Hainan Research Institute, Shanghai Jiao Tong University}

\author{Qichao Chang}
\affiliation{State Key Laboratory of Dark Matter Physics, Tsung-Dao Lee Institute, Shanghai Jiao Tong University, Shanghai 201210, China}

\author{Shuhua Hao}
\affiliation{State Key Laboratory of Dark Matter Physics, Tsung-Dao Lee Institute, Shanghai Jiao Tong University, Shanghai 201210, China}

\author{Jingtao Huang}
\affiliation{State Key Laboratory of Dark Matter Physics, Tsung-Dao Lee Institute, Shanghai Jiao Tong University, Shanghai 201210, China}

\author{Weilun Huang}
\affiliation{State Key Laboratory of Dark Matter Physics, Tsung-Dao Lee Institute, Shanghai Jiao Tong University, Shanghai 201210, China}

\author{Yifan Jiang}
\affiliation{State Key Laboratory of Dark Matter Physics, Tsung-Dao Lee Institute, Shanghai Jiao Tong University, Shanghai 201210, China}

\author{Shasha Liu}
\affiliation{State Key Laboratory of Dark Matter Physics, Tsung-Dao Lee Institute, Shanghai Jiao Tong University, Shanghai 201210, China}

\author{Hualin Mei}
\affiliation{State Key Laboratory of Dark Matter Physics, Tsung-Dao Lee Institute, Shanghai Jiao Tong University, Shanghai 201210, China}
\affiliation{Key Laboratory for Particle Astrophysics and Cosmology (MOE) \& Shanghai Key Laboratory for Particle Physics and Cosmology, Shanghai Jiao Tong University, Shanghai 200240, China}
\affiliation{Hainan Research Institute, Shanghai Jiao Tong University}

\author{Hengbin Shao}
\affiliation{State Key Laboratory of Dark Matter Physics, Tsung-Dao Lee Institute, Shanghai Jiao Tong University, Shanghai 201210, China}

\author{Yingwei Wang}
\affiliation{State Key Laboratory of Dark Matter Physics, Tsung-Dao Lee Institute, Shanghai Jiao Tong University, Shanghai 201210, China}

\author{Peilian Wu}
\affiliation{State Key Laboratory of Dark Matter Physics, Tsung-Dao Lee Institute, Shanghai Jiao Tong University, Shanghai 201210, China}

\author{Xin Xiang}
\affiliation{State Key Laboratory of Dark Matter Physics, Tsung-Dao Lee Institute, Shanghai Jiao Tong University, Shanghai 201210, China}
\affiliation{Key Laboratory for Particle Astrophysics and Cosmology (MOE) \& Shanghai Key Laboratory for Particle Physics and Cosmology, Shanghai Jiao Tong University, Shanghai 200240, China}
\affiliation{Hainan Research Institute, Shanghai Jiao Tong University}

\author{Fuyudi Zhang}
\affiliation{State Key Laboratory of Dark Matter Physics, Tsung-Dao Lee Institute, Shanghai Jiao Tong University, Shanghai 201210, China}

\author{Haoyan Zhang}
\affiliation{State Key Laboratory of Dark Matter Physics, Tsung-Dao Lee Institute, Shanghai Jiao Tong University, Shanghai 201210, China}

\author{Zhaozhi Zhang}
\affiliation{State Key Laboratory of Dark Matter Physics, Tsung-Dao Lee Institute, Shanghai Jiao Tong University, Shanghai 201210, China}

\author{Tailin Zhu}
\affiliation{State Key Laboratory of Dark Matter Physics, Tsung-Dao Lee Institute, Shanghai Jiao Tong University, Shanghai 201210, China}

\author{Yiwei Zhu}
\affiliation{State Key Laboratory of Dark Matter Physics, Tsung-Dao Lee Institute, Shanghai Jiao Tong University, Shanghai 201210, China}

\begin{abstract}

Large-volume neutrino telescopes such as IceCube and KM3NeT exploit natural ice or water as detection media, but their sparse instrumentation makes MeV neutrino interactions difficult to identify individually. We show that chlorine naturally abundant in seawater provides an intrinsic neutron-capture target that enables efficient tagging of electron antineutrinos through inverse beta decay. In contrast to the 2.2 MeV $\gamma$-ray from neutron capture on hydrogen, capture on chlorine produces an $\sim$8.6 MeV $\gamma$-ray cascade on a shorter time scale, yielding a distinctive delayed-coincidence signature. Using measured seawater properties at the TRIDENT site and a detector-level simulation, we find that chlorine-assisted tagging can extend the core-collapse supernova 5$\sigma$ discovery reach from $\sim$50 to over 100 kpc. We validate the underlying detection principle with a salted-water experiment at the China Jinping Underground Laboratory, observing correlated neutron-capture signatures under suppressed cosmic muon background conditions. Natural seawater thus transforms the deep ocean into an intrinsically enhanced antineutrino detection medium, opening new opportunities for MeV neutrino astronomy and distributed measurements of reactor and geophysical antineutrinos.

\end{abstract}

\maketitle

{\bf Neutrinos in the Deep Sea ---}
Large-volume neutrino telescopes have transformed our view of the high-energy Universe, detecting astrophysical neutrinos across the TeV--PeV energy range and beyond~\cite{IceCube_Diffuse, IceCube_TXS, IceCube_NGC1068, KM3-230213A}. 
Instrumenting cubic kilometres of transparent natural media with thousands of optical modules, these detectors provide enormous target masses for neutrino detection. Advances in optical module sensitivity now provide a complementary opportunity at much lower energies, where each module can act as an individual low-energy detector. Collectively, these modules offer a unique capability to search for MeV-scale neutrino interactions within an otherwise sparsely instrumented high-energy neutrino observatory. As we show below, the natural composition of seawater provides an additional, previously unexploited advantage for this purpose. 

At MeV energies, neutrinos provide direct information about both astrophysical and terrestrial phenomena. A nearby core-collapse supernova (CCSN) releases $\sim10^{58}$ neutrinos, providing a direct probe of the stellar core and explosion mechanism~\cite{SN1987A, SN1987A_rev, nu_diagnos_CCSN, SN_nu, Qian_2024}. Geoneutrinos provide a unique probe of the Earth's interior and heat production~\cite{geo_nu_Borexino_first, geo_nu_Borexino, geo_nu_KamLAND, geo_nu_JUNO, OBD_geo_nu}, while reactor antineutrinos enable precision neutrino measurements and remote reactor monitoring~\cite{Hanohano_2008,JUNO_TAO}. Improved sensitivity to these signals would unlock a new window on transient and terrestrial neutrino physics across the emerging global network of next-generation neutrino telescopes.

Electron antineutrinos are well suited for detection at these energies through inverse beta decay (IBD),
\begin{equation}
\bar{\nu}_e+p\rightarrow e^+ + n .
\end{equation}
The prompt positron signal followed by delayed neutron capture provides a powerful coincidence signature that has underpinned antineutrino detection since the first observation of the neutrino~\cite{Reines_Cowan}. For example, CCSN antineutrinos produce positrons with energies of order 10~MeV, while neutron capture on hydrogen in pure water produces a 2.2~MeV gamma ray after a characteristic time of $\sim$$200~\mu$s. Although such neutron tagging has been achieved in densely instrumented water Cherenkov detectors such as Super-Kamiokande and SNO+~\cite{Super-K_2008,SNO_neutron_proton}, the sparse instrumentation and high background rates of large-volume neutrino telescopes make this delayed signal extremely difficult to identify. Backgrounds include atmospheric muons, photomultiplier dark noise, and natural radioactivity, with $^{40}$K particularly important in seawater~\cite{ANTARES_K40,KM3Net_K40}. Consequently, existing low-energy searches with large-volume neutrino telescopes primarily rely on collective increases in counting rates rather than identifying individual neutrino interactions~\cite{Halzen_1995, IceCube_2011, Gen2_2021, IceCube_2023, KM3NeT_2021, KM3NeT_2021_DAQ, KM3NeT_2025}.

Fortunately, seawater itself provides a solution. Naturally occurring $^{35}$Cl ($\sim$$0.25\%$ number fraction) has a neutron capture cross section around 130 times larger than hydrogen, producing an $\sim$$8.6$~MeV gamma cascade with a shorter capture time of $\sim$$130~\mu$s. The higher-energy cascade produces substantially more detectable photons, while the reduced capture time narrows the coincidence window, turning seawater itself into an intrinsically favorable medium for event-level antineutrino identification without artificial doping.

In this work, we further develop our previously presented approach~\cite{TRIDENT_CCSN} to electron antineutrino detection in deep-sea Cherenkov detectors, exploiting neutron capture on naturally occurring chlorine in seawater, and provide its first experimental validation.

Using TRIDENT as an example detector~\cite{NA}, we demonstrate event-level identification of IBD interactions through detailed detector simulations incorporating realistic detector properties and measured seawater conditions at the deployment site, and evaluate the resulting boost to the detection horizon of core-collapse supernovae. We further validate the method's feasibility through a dedicated laboratory experiment using a multi-PMT optical module in salted water.

{\bf Antineutrino Identification in Seawater ---}
The performance of the proposed antineutrino detection method is evaluated using TRIDENT as an example detector~\cite{NA}. TRIDENT is a next-generation deep-sea neutrino telescope planned in the South China Sea, expected to instrument a volume of approximately 10~km$^3$ using $\sim1000$ detection strings. The primary detection units, known as hybrid digital optical modules (hDOMs)~\cite{hDOM_ICRC_2023}, consist of 31 $3''$ photomultiplier tubes (PMTs) arranged in a multi-PMT configuration. The pixelized design of the hDOM enables coincidence-based event identification at the module level, which is essential for the antineutrino detection method described in this work.

\textit{Backgrounds in the deep sea} -- Several sources contribute to the relevant hDOM background. First, radioactive decays generate optical signals both in seawater and in detector materials. Of particular importance is $^{40}\mathrm{K}$, whose $\beta$ decay (Q=1.3~$\mathrm{MeV}$) generally illuminates only a small number of PMTs because of its low energy. Site-specific measurements give a $^{40}\mathrm{K}$ activity of 10.78 Bq/kg in seawater, with U/Th-chain activities orders of magnitude lower~\cite{NA}; an activity of 61.0 Bq/kg is assumed for $^{40}\mathrm{K}$ in the hDOM glass~\cite{Gen2_2021}. Second, PMT dark noise generates uncorrelated photon-like hits that can form accidental coincidences; a dark-count rate of 1~$\mathrm{kHz}$ per PMT is assumed. Third, atmospheric muons remain an important correlated background despite the $\sim$3~km water overburden. At the TRIDENT depth, the residual muon flux is approximately 3$\times10^{-4}~\mathrm{Hz\,m^{-2}}$~\cite{KM3NeT_atm_muon}. We simulate the muon flux and energy spectrum with MUPAGE~\cite{MUPAGE} and obtain the single-hDOM response by averaging over detector depth. Unlike MeV interactions, however, throughgoing muons typically illuminate multiple hDOMs within a short time interval, providing an efficient inter-module veto. Bioluminescence is an additional deep-sea optical background~\cite{ANTARES_bioluminescence}, but it is strongly time-dependent and is typically localized rather than producing a detector-wide correlated excess like a CCSN burst. We do not model its time-dependent contribution in the present sensitivity estimate.

\begin{figure}
  \centering
  \includegraphics[width=1.0\linewidth]{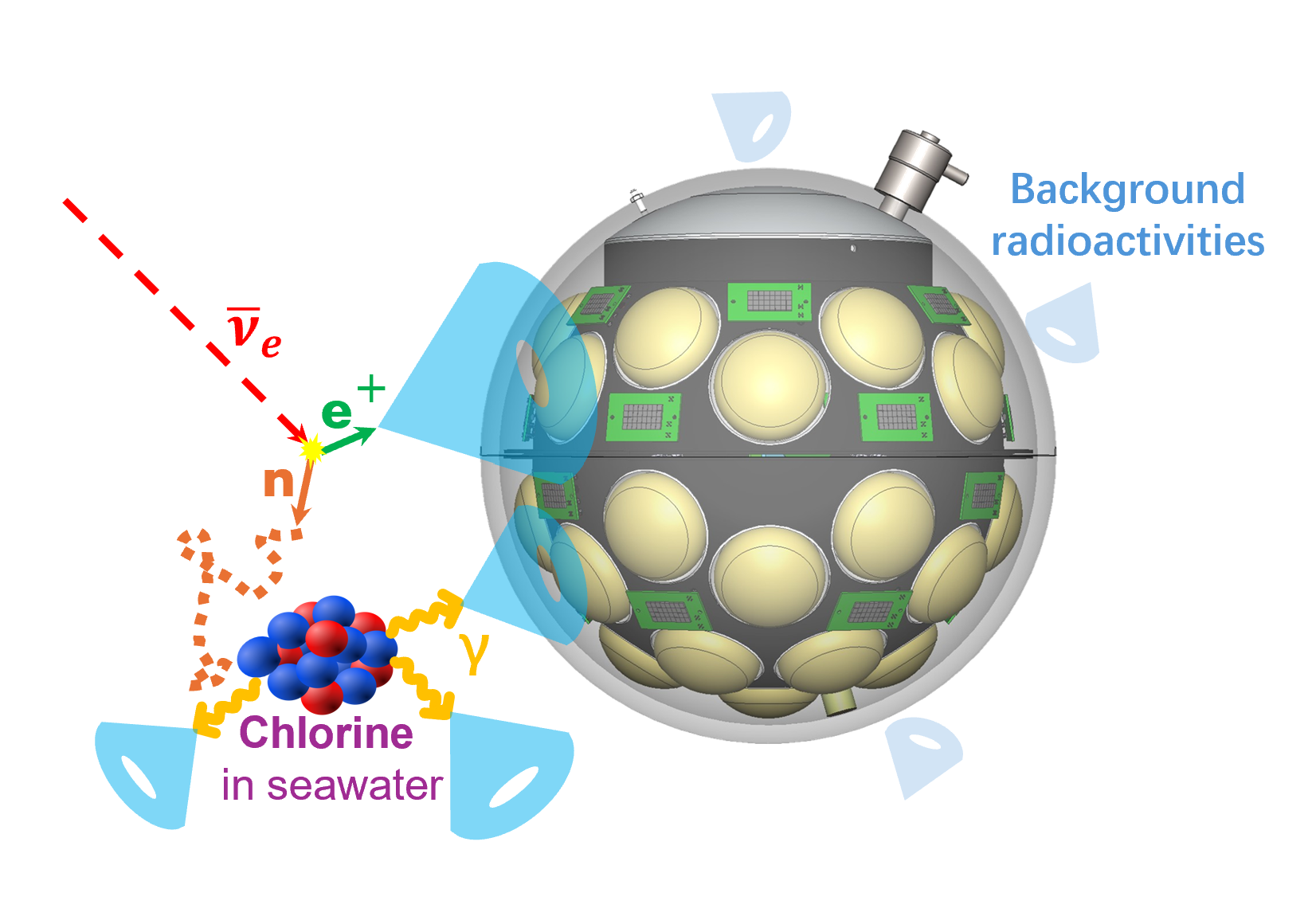}
  \caption{Schematic of IBD event-selection strategy in a seawater-based neutrino telescope. }
  \label{fig:filters_schematics}
\end{figure}

\textit{Identifying the prompt-delayed IBD signature} -- We exploit the characteristic temporal and spatial correlations between the prompt positron and delayed neutron-capture signals to distinguish IBD interactions from these backgrounds. The selection, illustrated schematically in Fig.~\ref{fig:filters_schematics}, combines four complementary observables: the local PMT coincidence level, the prompt-delayed time correlation, the angular distribution of detected photons, and an inter-hDOM muon veto.

To suppress the high rate of single-hit background noise on each hDOM, a trigger requires multiple PMTs within an hDOM to register photon hits within a narrow time window. We first define the coincidence level ($\mathrm{CL}$), as the number of distinct PMTs within an hDOM registering at least one photon hit in a 20~ns window~\cite{Trigger_TRIDENT}. Random dark counts and low-energy $^{40}\mathrm{K}$ decays fall steeply with increasing CL, whereas IBD positrons, with characteristic energies of order 10~MeV, produce higher local photon multiplicities. Atmospheric muons occur at much lower rates but generally yield still larger CL because of their high energies. A minimum CL requirement therefore provides a local measure of detected light yield and efficiently suppresses radioactivity and dark-noise backgrounds.

\begin{figure*}
  \centering
\centerline{\includegraphics[width=0.48\linewidth]{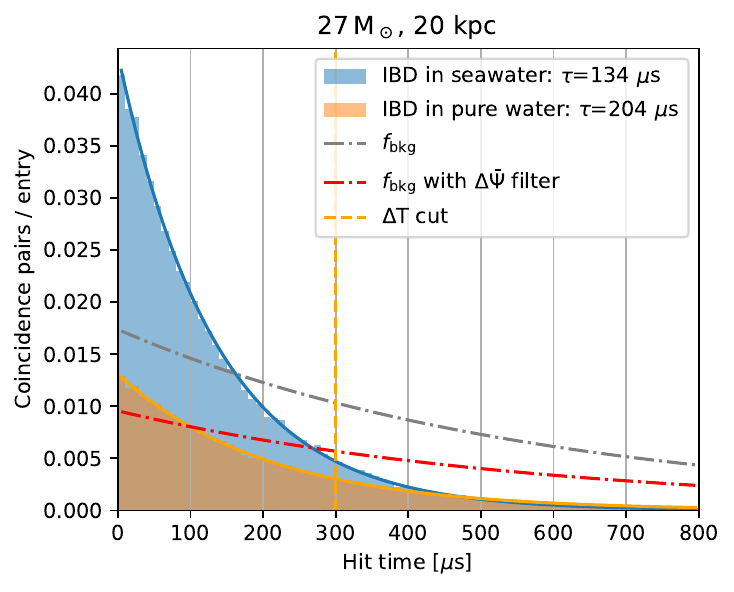}\hspace{0.03\linewidth}\includegraphics[width=0.45\linewidth]{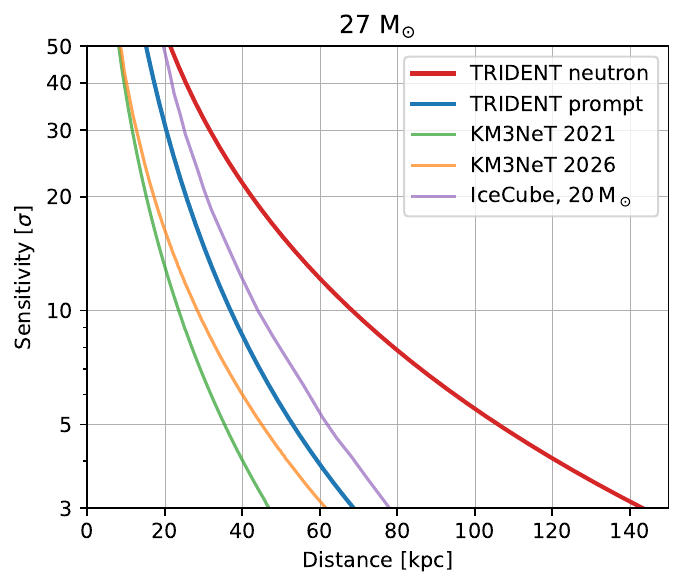}}
  \caption{\textit{Left}: Expected prompt--delayed time-difference distributions for IBD candidates per hDOM after applying the selection criteria described in the text, including both IBD events and accidental background pairs. The signal is normalized to the $27\,\mathrm{M_\odot}$ Garching model~\cite{Garching} at a distance of 20~kpc; the pure-water case is shown for comparison. \textit{Right}: Expected CCSN discovery significance as a function of distance for the prompt-only and prompt-delayed selections, compared with published sensitivities of IceCube for the $20\,\mathrm{M_\odot}$ Lawrence--Livermore model~\cite{IceCube_2011, Lawrence-Livermore} and of KM3NeT for the $27\,\mathrm{M_\odot}$ Garching model~\cite{KM3NeT_2021, KM3NeT_2025}, based on non-oscillated CCSN fluxes. }
  \label{fig:Sensitivity_compare}
\end{figure*}

The key additional handle is provided by the neutron. In pure water, neutron capture occurs predominantly on hydrogen,
$n+p\rightarrow d+\gamma\;(2.2~\mathrm{MeV})$,
making the delayed signal challenging to detect with sparse optical instrumentation. Seawater naturally contains $^{35}\mathrm{Cl}$, with a number fraction of 0.25\% compared with 66.1\% for hydrogen. Its thermal-neutron capture cross section, 43.6~barns, is approximately 130 times the 0.333~barn cross section of hydrogen. Capture on $^{35}\mathrm{Cl}$,
${}^{35}\mathrm{Cl}+n\rightarrow{}^{36}\mathrm{Cl}+\gamma\mathrm{s}$,
releases a $\gamma$-ray cascade with a total energy of approximately 8.6$~\mathrm{MeV}$. Together with an effective neutron-capture time of $\sim$$130\,\mu\mathrm{s}$ in seawater, compared with $\sim$$200\,\mu\mathrm{s}$ in pure water, the substantially larger light yield makes the delayed signal accessible to high-sensitivity optical modules, while the shorter time scale permits a narrower coincidence window to reduce accidental coincidences. The seawater simulation implemented in Geant4~\cite{GEANT4:2002, Geant4_neutron_HP} includes the relevant constituent nuclides, accounting for neutron capture on other nuclei in addition to chlorine; their abundances and neutron-capture properties are summarized in supplemental table~\ref{tab:neutron_capture}. The left panel of Fig.~\ref{fig:Sensitivity_compare} compares the simulated prompt--delayed time-difference distributions for IBD candidates in seawater and pure water, illustrating both the enhanced detectability and the shorter neutron-capture time in seawater.

The multi-PMT architecture provides an independent topological discriminator. For PMTs at angular coordinates ($\theta_1$,$\phi_1$) and ($\theta_2$,$\phi_2$), we define their angular separation $\Psi$ by
\begin{equation}
    \cos\Psi=\sin\theta_1\sin\theta_2\cos(\phi_1-\phi_2)+\cos\theta_1\cos\theta_2
    \label{eq:AS}
\end{equation}
For an event illuminating multiple PMTs, the mean angular spread $\bar{\Psi}$ is calculated over all hit-PMT pairs. Detected IBD interactions preferentially occur close to the triggered hDOM and therefore yield a more localized angular pattern, with $\bar{\Psi}$ concentrated between $40^\circ$ and $100^\circ$, whereas atmospheric-muon events span approximately $70^\circ$--$110^\circ$. For prompt candidates with $\mathrm{CL}_{\rm prompt}\ge5$, requiring $\bar{\Psi}<80^\circ$ rejects approximately 70\% of atmospheric-muon hits while retaining more than 80\% of IBD positrons.

The positron energy deposition and subsequent neutron capture are also approximately collocated on the scale relevant to an hDOM, despite the neutron traveling tens of centimeters while thermalizing. We define the mean hit direction for the prompt and delayed signals from the average of their respective hit-PMT orientation vectors. The prompt--delayed angular consistency variable, $\Delta\bar{\Psi}$, is then defined as the angle between these two mean directions. Requiring $\Delta\bar{\Psi}<100^\circ$ suppresses approximately 40\% of isotropically distributed $^{40}\mathrm{K}$ coincidences while retaining more than 90\% of IBD events.

Finally, the extended topology of atmospheric-muon tracks provides an efficient array-level veto. An hDOM candidate is rejected when another hDOM within 120~m registers $\mathrm{CL}\ge2$ within $\Delta T_{\rm veto}=150~\mathrm{ns}$. This requirement suppresses more than 80\% of atmospheric-muon events with $\mathrm{CL}\ge4$, increasing to more than 90\% for $\mathrm{CL}\ge14$, while reducing the IBD efficiency by less than 1\%.

\textit{Isolating Antineutrinos from Background} -- Prompt candidates are required to satisfy a minimum coincidence level of $\mathrm{CL}_{\rm prompt} \geq 9$ and an average angular spread $\bar{\Psi} < 80^\circ$. Delayed neutron candidates must satisfy $\mathrm{CL}_{\rm late} \geq 2$ and occur within a time window of $\Delta T < 0.3$~ms following the prompt event. To further suppress accidental coincidences, a prompt–delayed angular consistency requirement of $\Delta\bar{\Psi} < 100^\circ$ is applied. Finally, atmospheric muon backgrounds are reduced using an inter-hDOM veto, where events are rejected if two neighboring hDOMs within 120~m register $\mathrm{CL}\geq 2$ within a time window of $\Delta T_{\rm veto} $$=$$150$~ns. The final IBD selection criteria are summarized in supplemental table~\ref{tab:parameters}.

Following the application of all analysis cuts, the left panel of Fig.~\ref{fig:Sensitivity_compare} shows the resulting prompt-delayed time distribution per hDOM for a CCSN at 20$~\mathrm{kpc}$, assuming a $27\,\mathrm{M_\odot}$ progenitor~\cite{Garching}. Simulated signal IBDs were mixed with relevant backgrounds at their expected rates, where false IBD candidate pairs may be formed through random coincidences. The accidental background pairs exhibit a broader approximately exponential time-difference distribution than the correlated neutron-capture component, as shown in the figure. The red and gray dash-dotted lines represent the impact of the $\Delta\bar{\Psi}$ cut. The combination of local photon multiplicity, chlorine-enhanced neutron capture, prompt-delayed topology, and array-level muon rejection thus enables individual IBD interactions to emerge from the otherwise high-rate deep-sea optical background.

{\bf Extending the CCSN Discovery Reach ---}
Equipped with the IBD signal filters for each hDOM, we now evaluate the performance of the entire TRIDENT detector for CCSN antineutrino burst detection. The corresponding final signal efficiencies, background rates, and signal yields for the benchmark CCSN model are summarized in table~\ref{tab:parameters} of the Supplementary Material.

The median discovery significance is estimated using the Asimov expression~\cite{Sensitivity_formula}:
\begin{equation}
    Z=\sqrt{2\left[ (n_s+n_b)\ln\left(1+\frac{n_s}{n_b}\right)-n_s \right ]},
\end{equation}
where $n_s$ and $n_b$ represent the expected number of signal and background events, respectively. Here, we evaluate $n_s$ and $n_b$ in a 1~s time window, approximately corresponding to the duration of the electron-antineutrino-dominated accretion phase. The predicted sensitivity depends more strongly on the assumed CCSN progenitor model than the neutrino-oscillation scenario; we therefore adopt the non-oscillated $27\,\mathrm{M}_\odot$ Garching flux~\cite{Garching} as a common benchmark for comparison with other experiments.

The right panel of Fig.~\ref{fig:Sensitivity_compare} shows the resulting discovery significance as a function of CCSN distance, together with the incremental improvements from delayed-neutron tagging. Using the prompt events alone, TRIDENT is expected to achieve $5\sigma$ sensitivity to a $27\,\mathrm{M_\odot}$ CCSN out to approximately 50~kpc. Incorporating the prompt-delayed coincidence enabled by chlorine-enhanced neutron capture extends the 5$\sigma$ discovery reach to over 100~kpc, doubling the discovery reach. The same trend is also observed  for the $30\,\mathrm{M}_\odot$ Nakazato model~\cite{Nakazato_2012}, where the $5\sigma$ discovery horizon increases from approximately 60 to 125~kpc. This substantial gain illustrates how the enhanced neutron-capture signal naturally provided by seawater can be converted, through high-sensitivity photon detection and coincidence tagging, into improved astrophysical sensitivity.

{\bf Experimental Demonstration in Salted Water ---}
To demonstrate the feasibility of coincidence-based antineutrino detection at the single-module level, we performed a laboratory measurement using a TRIDENT hDOM at the China Jinping Underground Laboratory (CJPL). Its 2.4~km rock overburden suppresses the cosmic-muon flux to  $\sim$$\,0.2\,\mathrm{m^{-2}d^{-1}}$~\cite{CJPL_muon}, providing a controlled environment in which correlated neutron-capture signals can be isolated from optical backgrounds.

\begin{figure*}
    \centering
    \centerline{\includegraphics[width=0.9\linewidth]{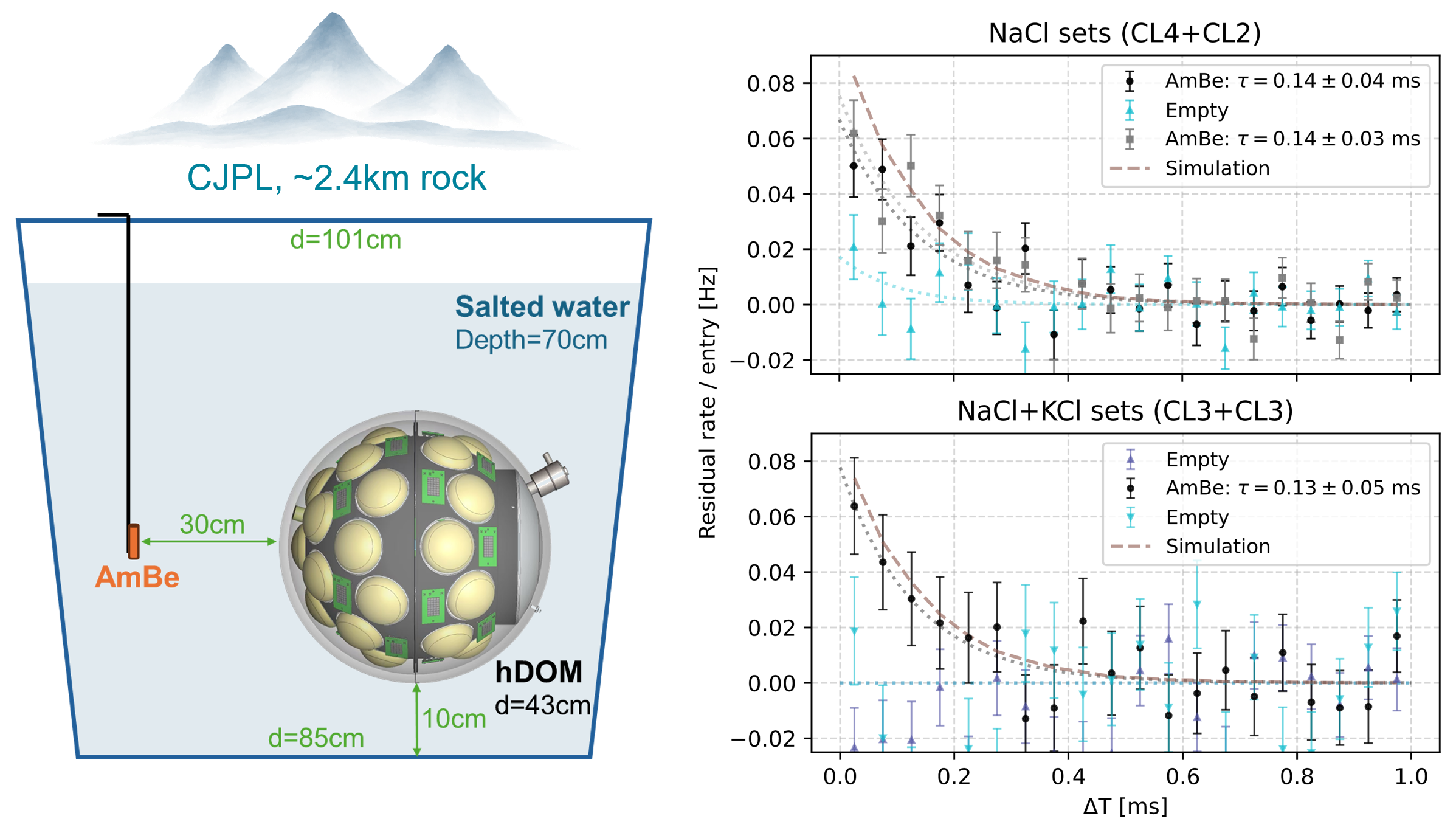}}
    \caption{\textit{Left}: Experimental setup at CJPL, showing the hDOM and AmBe neutron source deployed in a 0.5~ton water tank. \textit{Right}: Measured prompt--delayed time-difference ($\Delta$T) distributions for different water configurations after subtraction of the accidental background estimated from the $\Delta \mathrm{T}>0.5$~ms region. The gray dashed curve shows the simulated neutron capture time distribution. The different datasets correspond to independent hDOM measurement runs under source-on and source-off conditions.}
    \label{fig:AmBe}
\end{figure*}

A radioactive AmBe neutron source with an effective neutron activity of approximately 10~Bq was placed $\sim\,$30~cm from the hDOM in a water tank, as shown in the left panel of Fig.~\ref{fig:AmBe}. Neutrons are produced through the $^{9}\mathrm{Be}(\alpha,n)^{12}\mathrm{C}$ reaction, with a substantial fraction of the reactions populating the first excited state of $^{12}\mathrm{C}$ and producing an accompanying 4.4~MeV $\gamma$ ray upon de-excitation~\cite{AmBe_2023}:
\begin{align}
    \mathrm{^{241}Am} &\rightarrow \mathrm{^{237}Np} + \alpha , \notag \\
    \mathrm{^{9}Be} + \alpha &\rightarrow \mathrm{^{12}C^*} + n,\\
    \mathrm{^{12}C^*} &\rightarrow \mathrm{^{12}C} + \gamma\;(4.4~\mathrm{MeV}). \notag 
\end{align}
For these events, the prompt $\gamma$ followed by neutron thermalization and capture reproduces the essential prompt--delayed temporal structure of an IBD event, while allowing the neutron-capture component to be studied under controlled conditions.

Measurements were performed in pure water, NaCl solution at seawater concentration, and NaCl solution with additional KCl. For each configuration, corresponding source-off data were acquired to characterize the backgrounds. The hDOM data comprise two source-on and one source-off runs for the NaCl configuration, and one source-on and two source-off runs for the NaCl+KCl configuration. Approximately 10 h of data were collected for each configuration. 

Background events arise from radioactivity, PMT dark noise, and residual muons. These backgrounds produce random coincidences that follow an exponential distribution in the prompt–delayed time difference.
The expected background time constant is $\tau_{\rm bkg}$$\sim$$1\,\mathrm{ms}$ for a background rate of $\sim$1 kHz at $\mathrm{CL}\geq2$, which is significantly longer than the neutron capture time observed in salted water.

The accompanying 4.4~MeV $\gamma$ ray from the AmBe source provides a lower-energy prompt signal than typical CCSN positrons, and therefore a stringent test of local prompt--delayed coincidence detection. Consequently, reduced coincidence thresholds are used: $\mathrm{CL_{prompt}}=4$ and $\mathrm{CL_{late}}=2$ for NaCl, and $\mathrm{CL_{prompt}}=3$ and $\mathrm{CL_{late}}=3$ for NaCl+KCl. The higher delayed threshold for NaCl+KCl provides a purer neutron-capture sample in the presence of its higher radioactive background. To retain sufficient prompt statistics after imposing the higher delayed threshold, the prompt coincidence requirement is correspondingly lowered to $\mathrm{CL_{prompt}}=3$.

To isolate the neutron-capture signal, the long-time region ($\Delta T >0.5\,\mathrm{ms}$) was used to determine the accidental-coincidence background, which was then extrapolated to the full time-difference range. This 0.5~ms boundary was chosen to balance residual neutron-capture contamination against the statistical precision of the estimated background. Background-subtracted time-difference distributions are shown in the right panel of Fig.~\ref{fig:AmBe}, where an excess consistent with neutron capture is observed in the salted-water configurations.

We quantify the correlated excess with an extended binned-likelihood analysis. Two hypotheses are considered: a background-only hypothesis, $H_0$, and a signal-plus-background hypothesis, $H_1$, in which an exponential neutron-capture component is added to the background model. The likelihood function is defined as

\begin{equation}
\mathcal{L} = \prod_{i=1}^{N} \frac{e^{-\mu_i} \mu_i^{n_i}}{n_i!},
\end{equation}
where $n_i$ is the observed number of counts and $\mu_i$ is the expected count in the $i$-th bin. Under $H_1$, the expectation is modeled as

\begin{equation}
\mu_i(H_1) = \mu_{i, \mathrm{bkg}} + b + S \cdot \int_{\Delta T_i}^{\Delta T_{i+1}} \frac{1}{\tau_n} e^{-t/\tau_n} dt,
\end{equation}
where $\mu_{i,\mathrm{bkg}}$ is the extrapolated background contribution from the long-time fit, $b$ is a constant residual background, $S$ is the signal yield, and $\tau_n$ is the neutron-capture time constant.

The signal-plus-background hypothesis is compared with the background-only hypothesis using the likelihood-ratio test statistic
\begin{equation}
\mathrm{TS} = -2 \ln \left( \frac{\mathcal{L}(H_0)}{\mathcal{L}(H_1)} \right)
= 2 \left[ \mathrm{NLL}(H_0) - \mathrm{NLL}(H_1) \right].
\end{equation}
The statistical significance is calibrated using background-only pseudo-experiments that repeat the background-estimation and likelihood-fitting procedure while keeping the CL thresholds fixed. The resulting $p$-value is converted to a one-sided Gaussian-equivalent significance. 

Excesses at small time differences, with decay constants consistent with neutron capture, are observed in both the NaCl and NaCl+KCl configurations. For the NaCl dataset recorded with the AmBe source, the fitted excess has a significance exceeding $3.4\sigma$, with a best-fit neutron-capture time of $\tau_n = 0.14 \pm 0.04\,\mathrm{ms}$, consistent with neutron capture in seawater. In contrast, the corresponding AmBe-off dataset shows no statistically significant excess. For the NaCl+KCl dataset, the fitted significance is approximately $2\sigma$, with the reduced significance attributed to the higher radioactive background and the correspondingly larger statistical uncertainties shown in Fig.~\ref{fig:AmBe}. 

These measurements demonstrate that a single high-sensitivity, multi-PMT optical module can resolve correlated neutron-capture signals in salted water despite substantial optical backgrounds. They thereby validate the essential delayed-neutron-tagging mechanism underlying the proposed IBD identification. Together with the detector-level simulation, the measurement supports the feasibility of exploiting the enhanced neutron-capture signal in seawater for event-level antineutrino identification in a deep-sea neutrino telescope.

{\bf Conclusion ---}
We have demonstrated a new route toward event-level electron-antineutrino identification in deep-sea Cherenkov detectors by combining advanced photon detection with neutron capture on chlorine naturally present in seawater. Using detector-level simulations of TRIDENT incorporating measured optical properties and environmental backgrounds at its deployment site, we find that chlorine-enhanced delayed-neutron tagging doubles the 5$\sigma$ CCSN discovery reach, from approximately 50$~\mathrm{kpc}$ using the prompt signal alone to over $100~\mathrm{kpc}$. We further validate the essential neutron-tagging mechanism with a dedicated experiment at CJPL, where a single multi-PMT hDOM resolves correlated neutron-capture signals in salted water under controlled low-background conditions. The neutron-capture excess remains identifiable after adding KCl to introduce a $^{40}$K background representative of seawater radioactivity.

The same principle can be explored in other operating and planned seawater neutrino telescopes equipped with high-sensitivity, multi-PMT optical modules, including KM3NeT~\cite{mDOM} and P-ONE~\cite{PONE_mDOM}. Delayed neutron signal from hadronic showers produced in high-energy neutrino interactions at these detectors~\cite{Cascade_echo} could be improved by Chlorine-enhanced neutron capture, providing an additional handle on particle identification in deep-sea neutrino telescopes. Advanced multi-PMT modules proposed for the IceCube Upgrade~\cite{IceCubeUpgrade_mDOM} and IceCube-Gen2~\cite{IceCubeGen2_mDOM} may likewise enhance event-level sensitivity to MeV neutrinos in ice. Although the chlorine-assisted channel is absent there, the substantially lower intrinsic radioactive background may improve access to the 2.2~MeV delayed signal from neutron capture on hydrogen. Together, these complementary detector environments raise the prospect of extending MeV neutrino capabilities across a global network of large-volume neutrino observatories. 

More broadly, the mechanism demonstrated here is not restricted to high-energy neutrino telescopes. Because chlorine is supplied naturally by seawater, the enhanced neutron-capture signature requires neither artificial doping nor modification of the detection medium. This opens the possibility of distributed seawater-based antineutrino detectors whose locations can be chosen for specific science targets, from transient astrophysical sources to reactor and geophysical antineutrinos. Together with advanced photon detection, the natural salinity of the ocean thus becomes an asset rather than a background, opening the deep sea as a new detection medium for MeV antineutrinos.

\begin{acknowledgments}

We thank Yong-Zhong Qian for discussions from the outset of this project through 2025. We thank Liang Li and Zhigang Xiao for their careful review and valuable comments. The authors thank the TRIDENT (Hai-Ling) Collaboration for providing the platform to make this study possible. The authors also thank the PandaX group for kindly hosting and supporting the neutron experiment conducted at CJPL. This work was supported by grants from National Natural Science Foundation of China (No. 12427809) and Science and Technology Commission of Shanghai Municipality (No. 24590780100).

\end{acknowledgments}

\vspace{5mm}

\bibliography{ref.bib}

\appendix

\makeatletter
\@addtoreset{table}{section}
\makeatother
\setcounter{table}{0}
\renewcommand{\thetable}{\thesection\arabic{table}}

\newpage
\clearpage

\onecolumngrid
\begin{center}  
 \large Supplementary Material for \\ \textit{Natural Chlorine Enhances Antineutrino Detection in the Deep Ocean}
\end{center}
\vspace*{0.2cm}



\begin{table}[h!]
    \centering
    \caption{Seawater constituents included in the neutron-capture simulation, showing their number fractions, total $\gamma$-ray energy released following neutron capture ($\Sigma E_{\gamma}$), and thermal-neutron capture cross sections ($\sigma_n$)~\cite{Geant4_neutron_HP}.} 
    \begin{tabular}{|c|c|c|c|}
    \hline
       Nuclei& Number fraction & $\Sigma E_{\gamma}$ [MeV] & $\sigma_n$ [barn] \\
       \hline
       $^1$H & 66.1\%& 2.22&0.333  \\
       \hline
       $^{16}$O& 33.2\%&4.14 &0.0002  \\
       \hline
       $^{23}$Na& 0.28\% &6.96 &0.52  \\
       \hline
       $^{35}$Cl& 0.25\% &8.58 &43.6  \\
       \hline
       $^{37}$Cl& 0.08\% &6.11 &0.43  \\
       \hline
       $^{24}$Mg& 0.03\% &7.33 &0.054  \\
        \hline
        $^{32}$S& 0.02\% &8.64 &0.52  \\
       \hline
        $^{39}$K& 0.006\% &7.80 &2.1  \\
        \hline
        $^{40}$Ca& 0.006\% &8.36 &0.43  \\
       \hline
    \end{tabular}
    \label{tab:neutron_capture}
\end{table}


\begin{table*}[h!]
    \centering
    \caption{\textit{Upper part}: Selection criteria used for hDOM-level IBD identification. \textit{Lower part}: Final IBD detection efficiency, residual background rate, and expected signal yield per hDOM for a $27\,M_\odot$ Garching CCSN model~\cite{Garching} at 10~kpc. The IBD detection efficiency is averaged over interactions within 1~m of the hDOM center.}
    \begin{tabular}{|c|c|c|}
   \hline
   Selection criterion & Prompt & Late \\
   \hline
   CL\textsubscript{low}: Lower CL limit & 9 & 2\\
   $\bar{\Psi}$\textsubscript{upper}: Maximum angular spread &80°& - \\
    \hline
   $\Delta$T: Maximum prompt-delayed time difference & \multicolumn{2}{c|}{0.3~ms} \\
   $\Delta\bar{\Psi}$: Maximum prompt-delayed angle  &\multicolumn{2}{c|}{100°}\\
   $\Delta$T\textsubscript{veto}: Time difference veto for nearby hDOMs with CL$\geqslant$2 & \multicolumn{2}{c|}{150~ns}\\
    \hline
    \hline
    Performance metric & Positron only & With neutron \\
    \hline
    IBD detection efficiency (averaged in R$=$1~m)
    & 9.72\% & 3.30\% \\
    Residual background rate per hDOM [Hz]
    & $4.3\times10^{-2}$ & $2.8\times10^{-4}$ \\
    Signal counts per hDOM in 1~s
    & 0.212 & 0.083 \\
    \hline
    \end{tabular}
    \label{tab:parameters}
\end{table*}

\end{document}